**Title: Teaching Problem Solving in Undergraduate Physics Courses: An Endorsement for Deliberate Practice**

Author list: Kelly Miller, Olivia Miller, Georgia Lawrence
*Department of Physics and Division of Engineering and Applied Sciences, Harvard University, 9 Oxford Street, Cambridge, Massachusetts 02138*

**Abstract:**

Developing expert-like problem-solving skills is a central goal of undergraduate physics education. In this study, we investigate the impact of teaching explicit problem-solving frameworks, combined with deliberate practice, on students' problem-solving approaches. Using multidimensional scaling to analyze students' decision-making patterns, we compare the similarity of students taught with these methods to physics experts and to students taught with traditional repeated practice. Our results show that students who received structured frameworks and targeted feedback through deliberate practice exhibited problem-solving behaviors significantly more aligned with those of experts. These findings suggest that pedagogies emphasizing explicit strategy instruction and feedback are more effective than rote repetition for fostering expertise. We recommend integration of these approaches into physics curricula to better support the development of skilled and adaptive problem solvers.

### I. Introduction

The importance of developing problem-solving skills during an undergraduate education in science, technology, engineering, and mathematics (STEM) is widely accepted [1-3]. Recent engineering graduates claim problem solving as one of the most valuable skills they use in their careers [4]. In a 2011 study of physics PhD's 10 years into their careers, the AIP reported that problem-solving was cited as the second most important factor related to their career success (second only to hard work) [5]. While many physics education research studies have been conducted on students' conceptual difficulties and pedagogies that effectively promote content mastery, fewer studies have focused on how to develop and especially measure students' ability to solve physics problems [6]. Several physics education researchers have raised the issue of the community placing too much emphasis on conceptual understanding, at the expense of problem-solving skill development [7].

The purpose of this study is to investigate the following three research questions pertaining to teaching and assessing problem solving skills in the undergraduate physics classroom:
1. How can students' problem-solving skills be effectively measured and evaluated using a stand-alone diagnostic instrument or survey?
2. What could such an instrument tell us about the difference between novice (student) versus expert problem solvers?
3. How do novice problem-solving skills differ between those explicitly taught a problem-solving framework and those not taught a framework?

### II. Background

#### A. Problem-Solving and Cognition

The Program for International Student Assessment (PISA) defines the problem-solving competency as "an individual's capacity to engage in cognitive processing to understand and resolve problem situations where a method of solution is not immediately obvious." Problem solving has also been defined as "the cognitive process of moving towards a goal when the path is uncertain" [8]. Similarly, Carl Weinman and Wendy Adams define it as "cognitive processing directed at achieving a goal when no solution method is obvious to the problem solver" [9]. All three definitions emphasize that for true problem solving to occur, the solver does not know, at least initially, how to solve the problem. This leads to an important distinction between "authentic problems" and "exercises". Typical textbook problems are "exercises" because to solve them, students are required to apply a known procedure. The designation of a task as an "authentic problem" is dependent on the solver - if they know how to solve it, it is, by definition, not an authentic problem.

All the above definitions emphasize that problem-solving is a cognitive process. Other definitions have stressed that cognitively, the problem solver must construct knowledge and organize behavioral processes to cope with difficulties and as a result, problem solving is closely linked to creativity [10]. Cognitively, problem solving involves behaviors associated with educational goals that are high up on Bloom's Taxonomy. "Authentic" problems require cognitive processes which appear high on Bloom's Taxonomy because they require students to design, construct, develop and formulate solutions that they have never seen before. Metacognition (awareness and understanding of one's own thought processes) is also considered to be an important cognitive skill needed to be an effective problem solver [8]. Managing time and direction, determining the next step, monitoring understanding, asking skeptical questions and reflecting on one's own learning process are all important behaviors associated with successful problem solving [8].

B. Teaching Problem Solving Skills

Given its importance in post-college career success, there is a trend in STEM pedagogy towards emphasizing problem-solving skills over content knowledge [11]. Several teaching strategies have been developed to help students become competent problem solvers. Many introductory physics courses teach students to follow a prescriptive problem-solving procedure based on frameworks of expert problem solving. George Polya, a Hungarian Mathematician, developed the first framework, a four-step problem solving procedure [12]. Hellers' Competent Problem Solver is a similar, five-step strategy based on Heller and Reif's (1984) prescriptive model of physics problem solving [13]. Mazur (2014) developed a four-step problem-solving strategy to help frame the process for introductory physics students [14]. The steps involved in each of these frameworks are outlined in table 2. All three strategies involve an initial step of understanding the problem, making assumptions and thinking about how best to represent the physics. They all also involve a final, metacognitive step which prompts students to reflect on the reasonableness of their answer. The pedagogical value of this final step is well founded in the literature which shows that metacognition is needed to be an effective problem-solver [8]. This is, by no means, a comprehensive list of frameworks but provides an overview of those most commonly used in the context of physics problem solving.

| Step | Polya's Framework (1945) | Heller's Framework (1984) | Mazur's Framework (2014) |
|---|---|---|---|
| 1 | **Understand the problem** -translate the situation and goals into fundamental concepts and decide reasonable idealizations and approximations needed | **Focus the problem** -develop a qualitative description of the problem -first visualize the problem by creating a sketch and then write down a statement of what you want to find | **Getting Started** -organize information given and understand what is asked in the problem -think about which principles apply and which simplifying assumptions are helpful |
| 2 | **Devise a plan** -apply specialized techniques and use specific concepts to connect the situation with the goal | **Describe the physics** -use qualitative understanding of the problem to prepare for the quantitative solution | **Devise Plan** -ascertain what information is needed to solve the problem -what physical relationships (or equations) can be applied? |
| 3 | **Carry out the plan** -follow through with the specialized plan developed in "devise a plan" | **Plan the solution** -translate the physics description into a set of equations which represent the problem | **Execute Plan** -follow the steps outlined in the previous sections -carry out any necessary mathematical operations |

| 4 | **Look back**<br>-ask the questions: can I check the result? can I derive the solution differently? | **Execute the plan**<br>-execute the planned solution<br>-use equations planned to find a numerical solution | **Evaluate Result**<br>-reflect on your work, examine the result, determine whether your answer is reasonable |
|---|---|---|---|
| 5 | | **Evaluate the plan**<br>-check work to see if it is properly stated, reasonable, and has answered the question asked | |

Table I: problem solving frameworks used to scaffold the process for students

### C. Deliberative Practice versus Repeated Practice

Deliberate practice and repeated practice are both pedagogical practices used to build skills, but they differ significantly in structure, feedback, and effectiveness. Deliberate practice is more targeted, feedback-driven, and focuses on breaking skills down into smaller 'sub-skills'. Repeated practice on the other hand, is often unstructured, focuses on repetition and feedback is infrequent or non-existent. Deliberative practice has been shown to be more effective for developing expertise than simple repeated practice [15]. Repeated practice has been shown to be effective for skill maintenance or basic improvement whereas deliberative practice is more effective for skill mastery and expertise, specifically of high-level skills, like problem-solving [16,17].

### D. Novices Versus Experts

Much of the work on physics problem-solving has studied the differences between novice and expert problem-solvers [8, 9, 18-22]. Many studies have shown that students often learn how to solve quantitative problems by plugging values into algorithmic equations and pattern matching. Consequently, they are not developing the necessary skills to transfer their understanding to unseen situations [23-28]. Other studies have found differences in both the knowledge structure and the problem-solving strategy typically used by experts compared to novices. Reif & Heller [28] found that the main difference between novices and experts was how they organized and used their knowledge in the context of solving a problem. Experts organize their knowledge in a structured, cohesive way and can activate these knowledge structures when they are needed. Novices typically do not have these knowledge structures; rather their knowledge consists of random facts and equations which are context specific and lacking in conceptual meaning [29,30]. Experts redescribe the problem and use qualitative arguments to plan solutions before describing the details of the problem from a mathematical perspective. Novices tend to rush to quickly string together various, miscellaneous mathematical equations [28].

Wieman & Adams [9] found that when solving authentic problems, experts spend more time analyzing, planning, and managing their own behavior than novices and generally, demonstrate a more holistic and systemic approach. Heller [8] found that experts usually follow specific steps when solving a problem: understanding the problem, determining the concepts, making the plan, solving the problem, and evaluating the outcome. Novices, on the other hand, first try to solve problems by using mathematical expressions. Expert problem solvers take more time to understand the problem and the concepts involved as well as to explore the relationship between concepts. Novice problem solvers cannot establish these relationships, especially when the problem is complex [8].

### E. Measuring Problem Solving

A range of methodologies are used to measure problem solving skill level. They range from interviews [31] to think-aloud protocols [32] to expert self-reflections [12], to observing participants solving problems in a lab setting [33], to computer-delivered assessment [34].

#### 1. Computer-delivered Assessments

Computer-delivered assessment is emerging as a common methodology for studying problem solving. The PISA 2012 assessment of problem solving was computer-delivered for the first time. These assessments involve problems that do not require any specific disciplinary knowledge for their solution so the focus can be placed on measuring the cognitive processes fundamental to problem solving. Students are presented with a scenario and a series of test items about the scenario are posed to measure students' performance in four cognitive processes involved in problem solving (exploring and understanding, representing and formulating, planning and executing, monitoring and reflecting) [34]. Each test item occupies a single computer screen, and students proceed from item to item without being able to skip items or go back to previous items. Test items consist of a variety of response formats which can be automatically coded (multiple-choice, drag-and-drop, constructed-response) and free text response which requires coding by experts.

*2. Interviews/ Think-aloud Protocols and Rubrics*

Many studies involving the measurement and assessment of problem-solving use interviews and/or think-aloud protocols. Students (and sometimes experts) are interviewed by researchers who pose questions designed to understand the approach being taken to solving a problem. Think-aloud is a semi-structured cognitive interviewing method in which a person is asked to verbalize their thought process as they do a specific task, during which they are recorded (on paper, audio or video) for further analysis. In the study published by Alii et al. (2016) [32], 21 students were asked to narrate their thinking while solving physics problems and all data were recorded. Afterwards, qualitative interviews were held with the students. Interviews (or think-aloud recordings) are typically transcribed, and the text is analyzed and categorized (or coded) by several experts in the subject-matter domain. Interrater reliability is monitored by measuring the level of agreement between the expert coders to ensure validity. The interviews and generation of codes is often iterative. Rubrics are a common measurement tool used to evaluate students' written work and are often used as part of the methodology in problem solving studies. Halim et al. (2016) studied students' ability to apply problem solving strategies in physics [35]. Students were asked to solve routine problems on paper and rubrics were used as the measurement tool. Burkholder et al (2020) developed a solution template for students to use as they solve problems [36]. Responses to students' problem solving using this template were collected and then the template was used as a rubric for measuring how expert-like students' problem-solving thinking was.

F. Characterization of Expert Problem-Solving Process

Recently, important research has been done using think-aloud interviews to characterize the sub-skills of problem-solving to understand how experts make decisions when they solve problems. Price et al (2021) interviewed 52 science and engineering experts and asked them to explain how they go about solving problems in their fields [31]. The study found a surprising degree of overlap in the decisions made by experts across various fields. This consistency suggests that there are common decision elements shared across disciplines in science and engineering. Table 1 summarizes this framework and provides a list of the decisions that Price *et al* found experts commonly made when solving problems in their fields.

| Category Name | Decision | Description |
|---|---|---|
| A: Selection and Goals of the Problem | 1. What is important in the field? | Identifying key issues or topics of significance in the field. |
| | 2. Opportunity fits solver's expertise? | Evaluating if the problem aligns with the researcher's knowledge and skills. |
| | 3. Goals, criteria, constraints? | Defining objectives, criteria for success, and limitations. |
| B: Important Features and Information | 4. Important features and info | Determining available information relevant to problem solving |
| | 5. What predictive framework? | Deciding among possible predictive frameworks |
| | 6. How to narrow down the problem? | Often involves formulating specific questions and hypotheses |
| | 7. Related problems? | Identifying similar problems and useful solutions from other contexts. |

|  | 8. Potential solutions? | What are potential solutions? (This is based on experience and fitting some criteria for solution they have for a problem having general key features identified.) |
|---|---|---|
|  | 9. Is the problem solvable? | Deciding if problem is solvable and if solution is worth pursuing given the difficulties, constraints and uncertainties |
| C: Predictions and Frameworks | 10. Approximations and simplifications to make? | Simplifying the problem to make it easier to solve. |
|  | 11. How to decompose into subproblems? | Breaking the problem up into more tractable subproblems |
|  | 12. Most difficult or uncertain areas? | Identifying areas of particular difficulty and/or uncertainty. |
|  | 13. What information is needed? | Identifying the information needed to solve the problem. |
|  | 14. Priorities | Prioritizing the most important considerations. |
|  | 15. Specific plan for getting information | Identifying general requirements for the problem-solving approach they will pursue. |
| D: Narrowing Down the Problem | 16. Which calculations and data analysis? | Identifying calculations and data analysis needed. |
|  | 17. How to represent and organize information? | Identifying the best way to represent and organize available information to provide clarity and insight. |
|  | 18. How believable is the information? | Is information valid, reliable and believable? |
|  | 19. How does information compare to predictions? | Comparing new information to expected results. |
|  | 20. Any significant anomalies? | Figuring out how to follow up on anomalous results. |
|  | 21. Appropriate conclusions? | Identifying appropriate conclusions based on the data. |
|  | 22. What is the best solution? | Deciding on the best solution(s). |
| E: Related Problems | 23. Assumptions and simplifications appropriate? | Deciding if previous simplifications and predictive frameworks are still appropriate. |
|  | 24. Additional knowledge needed? | Deciding if more information is needed. |
| | 25. How well is the problem-solving approach working? | Deciding if the problem-solving approach needs to be modified. |
| | 26. How good is the solution? | Deciding if the chosen solution is adequate. |
| F: Implications and Communication | 27. Broader implications? | Considering the wider impact and significance of the findings. |
| | 28. Audience for communication? | Identifying who needs to be informed about the findings. |
| | 29. Best way to present work? | Deciding how to effectively share the research outcomes. |
| G: Ongoing Skill and Knowledge Development | 30. Stay up to date in field | Continuously learning about the latest developments in the field. |
| | 31. Intuition and experience | Using accumulated intuition and experience in problem solving. |
| | 32. Interpersonal and teamwork | Developing skills necessary for effective collaboration. |
| | 33. Efficiency | Enhancing productivity and the ability to make timely decisions. |
| | 34. Attitude | Maintaining a positive and open-minded approach to challenges. |

Table II: Problem solving decision framework presented in Price *et al*, (2021) [31]

The framework of decisions defined by Price *et al* (outlined in table 1) provides a structured approach to assessing problem-solving skills and it is this framework that we base our analysis for this study. The rows shaded in grey, in table 2, represent the decisions that we focus on.

### III. Methods

We developed a computer-delivered physics problem solving diagnostic instrument designed to measure an individual's ability to solve a complex physics problem. This instrument consists of a series of questions which asks respondents to consider a problem and, without explicitly solving it, devise a procedure to solve it. The instrument was designed to decouple (as much as possible) a respondents' problem-solving skill-level from their mastery of the underlying physics. Respondents are evaluated on their ability to devise a plan for solving the problem as opposed to solving it. We used Price *et al*'s (2021) [31] problem solving decision framework (outlined in table 2) to code respondents' devised plans for solving the problem presented in the diagnostic instrument. We administered this survey to experts of physics and physics students (novices). We tested two different populations of students from two different introductory physics courses at Harvard University - each with a different pedagogical approach to problem solving. One course explicitly taught a problem-solving framework, and the other course did not.

#### A. Problem Solving Diagnostic Instrument

The problem-solving diagnostic instrument was designed to elicit the problem-solving process a respondent would take to solve a physics problem without having them solve it. The problem presented is in the context of mechanics and the stem is shown in figure 1.

A ship building company has just finished building a new steel ship and they are outfitting the interior. They need to mount flat screen monitors in many of the rooms. The traditional way to do this is to drill holes in the steel walls and bolt the monitors on. This takes a lot of labor drilling the holes in the steel walls, and it makes it very difficult to move the monitors to different locations as needs change. Someone in the company suggested the idea of bolting magnets to the back of the monitors and then holding them on the walls magnetically. They come to you as a physics consultant to figure out if this method would be practical. They also provide you with 4 magnets.

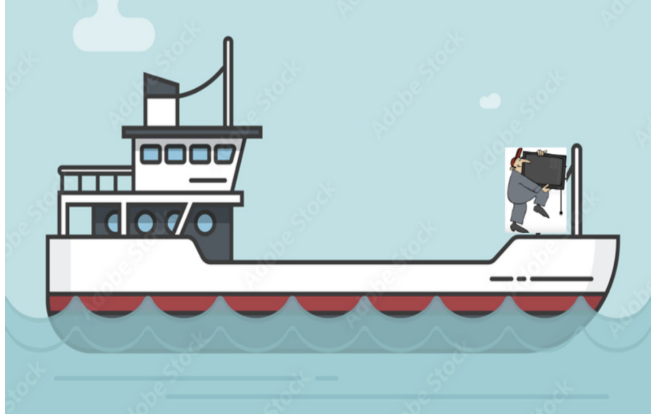

FIG 1. problem stem for the problem-solving diagnostic instrument

The context of the problem is that you are working for a shipbuilding company and have been assigned to determine whether you can attach a monitor to the wall of the ship using a set of four magnets. The diagnostic consists of 14 questions, and, like the PISA (2012) [34] problem-solving assessment, the questions consist of a variety of response formats; constructed-response, multiple-choice, many-choice, drag-and-drop. The diagnostic is designed to start with broad (un-scaffolded) questions about the process the respondent would use to solve the problem. The first question on the diagnostic is:

> "Imagine you are working for a physics consulting firm, and you are recommending the most secure way to hang the monitor on the wall of the ship (assuming you only have 4 magnets). Write a detailed plan of your recommendation. Include as many details as you think are necessary for someone else to be able to implement it. Your response should be 1-2 paragraphs long."

After committing to a detailed plan in the first question, respondents are then guided through 13 more questions probing key features of the problem and potential solution paths, but the questions are more targeted (e.g., what assumptions would you make, draw a diagram). The instrument starts with the broad, un-scaffolded "devise a plan" question so that it does not influence respondents' process-oriented thinking with the more scaffolded questions.
This progression from the open-ended to the scaffolded questions is an intentional structure designed to gather as much information as possible from respondents with a wide range of problem-solving skills. Increasing the scaffolding as the respondent progresses from one section to the next allows us to gather information about a respondents' problem-solving process even if they are not able to devise a cohesive plan in the first section.

The instrument is administered through the Harvard Qualtrics interface. Qualtrics is a web-based survey software for creating surveys. It also collects and stores data from survey respondents. Upon logging on to each survey on Qualtrics, respondents are told that they will be presented with an introductory physics problem with questions designed to probe their thinking about how to solve the problem. Respondents are also told that the questions will be presented in sections, and they will not be able to go back to previous sections once they advance to the next one. They are also instructed to only use the information that is provided in the problem, and to not use any outside resources. Finally, they are told that they should complete the diagnostic in one sitting and that while there is no time-limit, it should take around 45 minutes to complete.

B. Instrument Validation

We validated the instrument using the methodology described in "Development and Validation of Instruments to Measure Learning of Expert-Like Thinking" (Adams & Weiman, 2011) [37] which stresses the use of student interviews as the most important part of developing and validating formative assessments designed to measure the expertness of students' thinking. The steps taken to develop and validate the problem-solving diagnostic instrument are outlined below. The process was iterative and involved student and expert interviews at several stages.

1. Consulted several experts to solicit advice and feedback on the initial problem stem. These experts were all college instructors of introductory physics who had been teaching for at least 10 years. Several candidate problems were considered and iterated on. The following selection criteria were considered:
   - Authenticity - problems where the solution method is not immediately obvious to even expert problem solvers
   - Complexity - problems complex enough that, to solve them, the solver must state their assumptions and have a detailed method
   - Open-ended - problems with multiple solution paths which depend on initial assumptions
2. Conducted interviews of both students and experts. With the problem stem developed, we conducted six interviews of introductory physics students at Harvard University and four introductory physics instructors. We showed them the problem and asked them to talk through the ways in which they would solve them. These ten interviews allowed us to identify student thinking about the problem and the various ways this was different from expert thinking. We revised the problem stem for clarity.
3. Created open-ended survey questions to probe student thinking more broadly. This first version of the survey was administered to 18 introductory physics students at Harvard. We subsequently interviewed a subset of 6 of these students (chosen based on the diversity of their responses to the first "devise a plan" question).
4. Iterated the instrument based on students' responses and interview findings. We converted some of the open response questions into multiple-choice questions based on common student responses.
5. Conducted validation interviews on the new version of the survey with five novices and four experts (instructors of introductory physics but different from those interviewed in step 2).
6. Administered instrument to students and experts on a wider scale.

C. Coding the 'Devise a Plan' Responses

After the final version of the instrument was developed, we conducted an experiment to determine if physics students' responses to the first survey item (devise a plan) differed depending on whether they were explicitly taught a problem-solving approach in their introductory physics class. We administered the survey to two large introductory classes at Harvard; one of which taught students a problem-solving process ($N$=72) and the other which did not ($N$=117), as well as 13 experts. The experts were all college physics instructors who had been teaching introductory physics for at least 10 years. Three content experts studied each of the responses to the first survey item and coded it based on Price *et al's* [31] list of decisions made by experts when solving problems in their fields (outlined in table 1). Each of the three coders made a list of decisions from table 1 reflected in each of the 'devise a plan' responses. The decision codes were compared across the three coders and all cases where there was not perfect agreement were discussed and resolved.

Figure 2A and B show two examples of novice responses to item one on the survey and the corresponding decision codes

2A:

| | |
|---|---|
| <mark>Assuming that the monitor is not too heavy and the magnets are strong enough,</mark> the most secure way to hang the monitor on the wall of the ship <mark>would be to use all four magnets in a configuration known as a "four-point suspension system."</mark> In a <mark>four-point suspension system, two magnets are attached to the top corners of the monitor, and two more magnets are attached to the corresponding points on the wall.</mark> <mark>The magnets should be arranged so that the north pole of one magnet faces the south pole of the other magnet.</mark> By using this configuration, the <mark>weight of the monitor will be evenly distributed across all four magnets,</mark> ensuring that the monitor stays securely attached to the wall of the ship. Additionally, the four-point suspension system provides redundancy, meaning that if one magnet fails, the other three will still be able to support the weight of the monitor. | Approximations & simplifications? (10)<br><br>Potential solutions? (8)<br><br>**Related problems? (7)**<br><br>Important features & info? (4)<br><br>Predictive framework? (5)<br><br>Goals, criteria, constraints? (3) |

2B:

| | |
|---|---|
| <mark>I would want to find the strength of the magnets, especially through the distance of the walls to determine if they are strong enough to hold the weight of the monitor.</mark> <mark>Additionally, there would need to be some kind of stress test to see how the magnets would hold if the ship encounters rough waters.</mark> <mark>If the magnetic field is strong enough to hold the monitor in place past the limits of what is expected to be encountered, then it would seem like a viable option to hang the monitor.</mark> <mark>However, we would also need to take into consideration our ability to remove the magnets if we want to move the monitor, which is the entire reason we are looking at magnets in the first place, as well as any potential damage that could come to the monitor from being in close proximity to the magnet.</mark> | What info needed? (13)<br><br>Specific plan for getting info?(15)<br><br>Goals, criteria, constraints? (3)<br><br>**Most difficult or uncertain areas? (12)** |

FIGS 2A&B. two examples of novice responses to the 'devise a plan' question on the problem-solving diagnostic instrument. Color codes correspond to decision codes (from table 1)

### D. Testing Novices from two classes with different pedagogical approaches to problem solving

We administered the problem-solving diagnostic survey to two groups of novices in two different, introductory, calculus-based introductory physics classes at Harvard University. Both courses are comprised of students with similar physics backgrounds (non-majors) and historically, pre-course FCI scores from both courses have indicated that both populations come in with a similar baseline physics knowledge. While both courses covered the same content, one course (DP1) explicitly taught students to use a four-step problem solving procedure (Mazur's framework from table 1) for all problems encountered throughout the course. The other course (RP2) did not present or model a framework for problem solving. Both courses required students to solve the same number and types of problems (during class and in the homework), however their approach to teaching problem solving was different. At the beginning of DP1, the instructor outlined each of the four steps in Mazur's (2014) framework [14] (Getting Started, Devise a Plan, Execute the Plan, Evaluate the Result) and walked through several examples of problems solved using this procedure. This framework is also modeled throughout the required textbook used in every aspect of the course where students are required to solve a problem. Students in DP1 are required to explicitly use this framework for every problem they encounter in course, including any submitted homework problem. DP1 students receive consistent, targeted feedback on their problem-solving process within the prescribed framework. To receive credit for submitted homework, DP1 students must have each of the four steps enumerated for all problems. DP1 students are assessed on following the four-step problem solving process instead of the correctness of the final solution. In RP2, on the other hand, there is no problem-solving framework presented, and students are not provided feedback on their problem-solving process. The assessment for RP2 is based solely on correctness and not on following a specific problem-solving procedure. From a pedagogical point of view, DP1 follows deliberative practice, with an emphasis on breaking the problem-solving process into sub-skills and providing students with targeted feedback. RP2, on the other hand, uses a repeated practice (unstructured) approach, which focuses on repetition.

## IV. Results

Table 3 summarizes the frequency of the decisions found in all the responses to the "devise a plan" question on the problem-solving diagnostic instrument. Devising a plan only involves the first 3 categories of decisions from table 2 (defining goals of the problem, identifying important features and information, creating predictions and frameworks) and therefore, only decisions three through fifteen are represented in the responses. The frequency with which these decisions appear across all the responses ($N$=202) is found in table 3.

| Decision | Frequency of Decision in Responses (N=202) |
|---|---|
| 3. Goals, criteria, constraints? | 157 |
| 4. Important features and info | 134 |
| 5. What predictive framework? | 85 |
| 6. How to narrow down the problem? | 2 |
| 7. Related problems? | 5 |
| 8. Potential solutions? | 142 |
| 9. Is problem solvable? | 19 |
| 10. Approximations and simplifications to make? | 27 |
| 11. How to decompose into subproblems? | 7 |
| 12. Most difficult or uncertain areas? | 55 |
| 13. What information is needed? | 66 |
| 14. Priorities | 35 |
| 15. Specific plan for getting information | 47 |

Table III: frequency of decisions (from Price *et al*'s problem solving framework (table 2)) represented in the 'devise a plan' responses

A. Novices versus Experts

To determine what we can learn from this problem-solving diagnostic instrument about the differences between novice and expert problem solvers we compared the frequency with which novices made the decisions listed in table 3 to that of experts. This is represented in figure 3 which shows the percentage of each group (novices/experts) that demonstrated problem solving decisions (3 through 15) in their response to the 'devise a plan' question on the problem-solving diagnostic instrument. We performed a chi-squared analysis to determine which decisions showed statistically significant differences between novices and experts.

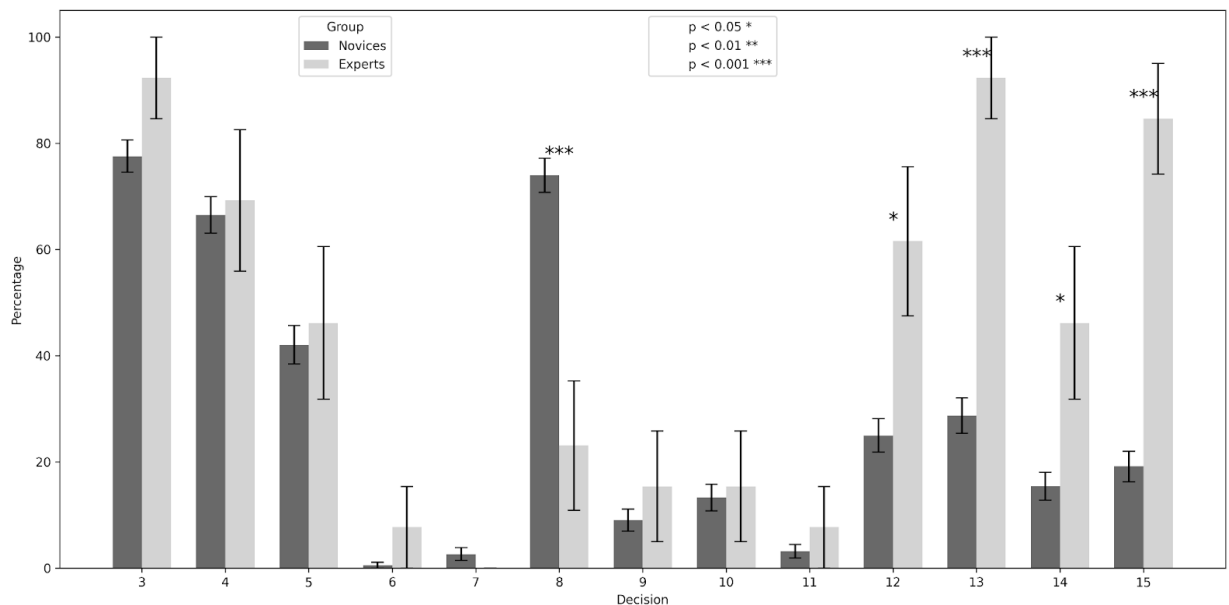

FIG 3. percentage of novices (dark grey) and experts (light grey) which showed evidence of problem-solving decisions (3 through 15) in their 'devise a plan' response. Error bars represent the standard error of the mean, asterixis denote the p-values for a chi-square test between the two groups.

Figure 3 shows that experts consider the most difficult or uncertain areas of the problem (decision 12), statistically significantly more than novices. Experts are also better at focusing on the information they will need to solve the problem (decision 13), the specific plan for getting information (decision 15), and how to prioritize the most important considerations (decision 14). Novices, on the other hand, are more likely than experts to focus on the potential solution (decision 8).

B. Comparing two different pedagogical approaches to problem solving

To determine if novice problem-solving skills differ between those explicitly taught a problem-solving framework and those not taught a framework, we compared the frequency with which DP1 students made the decisions listed in table 3 to that of RP2 students. DP1 explicitly taught students how to apply a framework to solving problems whereas RP2 did not. This comparison is represented in figure 4 which shows the percentage of each group (DP1/RP2) that demonstrated problem solving decisions (3 through 15) in their response to the 'devise a plan' question on the problem-solving diagnostic instrument. We performed a chi-squared analysis to determine which decisions showed statistically significant differences between novices and experts.

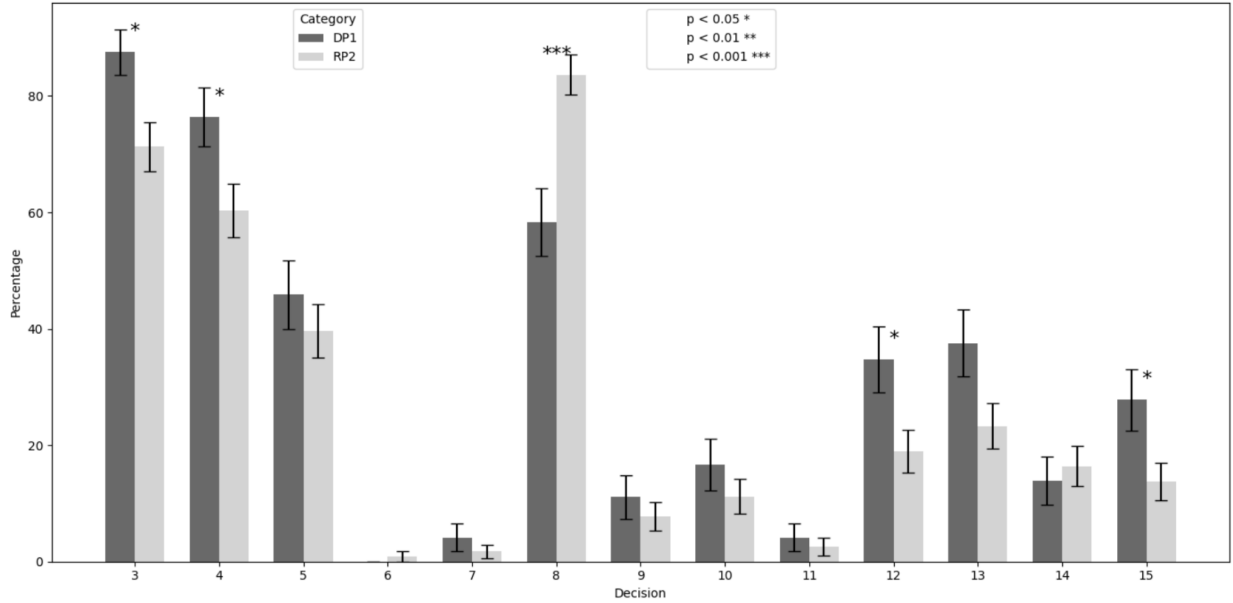

FIG 4. Percentage of DP1 students (dark grey) and RP2 students (light grey) which showed evidence of problem-solving decisions (3 through 15) in their 'devise a plan' response. Error bars represent the standard error of the mean, asterixis denote the p-values for a chi-square test between the two groups.

Figure 4 shows students taught a problem-solving procedure (i.e. DP1 students) define goals, criteria and constraints (decision 3), determine core features or ideas related to the problem (decision 4), identify areas of particular difficulty and/or uncertainty (decision 12) and outline a specific plan for getting information (decision 15), statistically significantly more than students not taught a problem-solving procedure (i.e. RP2 students) who are more likely to focus on a potential solution (decision 8).

| Decision | Novice (%) | Expert (%) | P-value (chi squared) | DP1 (%) | RP2 (%) | P-value (chi squared) |
|---|---|---|---|---|---|---|
| 3) Goals, criteria, constraints? | 77.1 | 92.3 | 0.3 | 87.5 | 70.7 | 0.01 |
| 4) Important features and info? | 66.5 | 69.2 | 1.0 | 76.4 | 60.3 | 0.03 |
| 8) Potential solution? | **73.9** | 23.1 | 0.0003 | 58.3 | **83.6** | 0.0002 |
| 12) Most difficult or uncertain areas? | 25.0 | **61.5** | 0.01 | **34.7** | 19.0 | 0.02 |
| 13) What information is needed? | 28.7 | **92.3** | 0.00001 | **34.5** | 23.3 | 0.05 |
| 14) Priorities? | 15.4 | 46.1 | 0.01 | 13.8 | 16.4 | 0.8 |
| 15) Specific plan for getting information | 19.1 | **84.6** | 0.00001 | **27.8** | 13.8 | 0.02 |

Table IV: summary of decisions with statistically significant differences between novices/expert groups and/or DP1/RP2 groups

Table 4 provides a summary of the statistically significant results from figures 3 and 4. To determine which group of students was more expert-like in their decisions, we bolded the larger percentage for decisions which showed significant differences between both experts and novices and between DP1 and RP2 students. For example, decision 8, the results from both chi-squared tests are significant. A higher percentage of novices (73.9%, compared to 23.1% of experts) referenced decision 8 in their response. Decision 8 (focusing on a potential solution) was also referenced more by RP2 students (compared to DP1 students) (83.6% compared to 58.3%), implying that RP2 students are less aligned with experts on this decision. The other three decisions which show statistically significant differences between both sets of groups (decisions 12, 13 and 15), all show a higher percentage of experts and a higher percentage of DP1 students. From this, it appears the pattern is the decisions more commonly made by experts are also more commonly made by DP1 students. Overall, our data shows that novices are more likely to focus on generating a potential solution (i.e. where to place the magnets on the back on the monitor) whereas more expert problem-solvers focus on what information is needed to solve a problem and a specific plan for *how* to solve it. Experts are also more likely to identify the difficult or uncertain areas of a problem (for example – the fact that the monitor is being mounted on a ship and the ship will rock back and forth).

To more rigorously determine which group of students is more expert-like in their problem-solving decisions, we used two common measures: Euclidean distance and Manhattan distance. To calculate the distances, we first summarized the pattern of responses from each group—experts, DP1 students, and RP2 students—by representing their decisions as a series of numerical values (a vector), where each number corresponded to a particular type of response (decision). Euclidean distance measures the straight-line distance between two groups' patterns of responses (like measuring the shortest distance between two points on a map), whereas Manhattan distance adds up the absolute differences in each response (like following a grid of city blocks from one location to another). By applying these distance measures, we were able to rigorously assess which student group's decisions were most like those of the experts. Smaller distances mean two groups responded more similarly, while larger distances indicate more substantial differences in their problem-solving approaches. The Euclidean distance between DP1 students and experts was 0.97, whereas the distance between RP2 students and experts was 1.3. This indicates that the pattern of responses in DP1 is closer to that in experts than RP2 is to experts. The Euclidean distance between DP1 and RP2 was 0.44, suggesting that the student groups are more alike compared to their individual distances from experts. Similar trends were observed with the Manhattan distances: The distance between DP1 and experts was 2.4. The distance between RP2 and experts was 3.3. The distance between DP1 and RP2 was 1.2. These results collectively suggest that DP1 students' response pattern is more like that of experts than RP2 students are, with DP1 students and RP2 students being the most like each other.

| **Euclidean distance** | **DP1** | **RP2** |
|---|---|---|
| Experts | 0.97 | 1.3 |
| DP1 | | 0.44 |
| **Manhattan Distance** | **DP1** | **RP2** |
| Experts | 2.4 | 3.3 |
| DP1 | | 1.2 |

Table V: Euclidean and Manhattan distances between the three categories of respondents (experts, DP1 students and RP2 students)

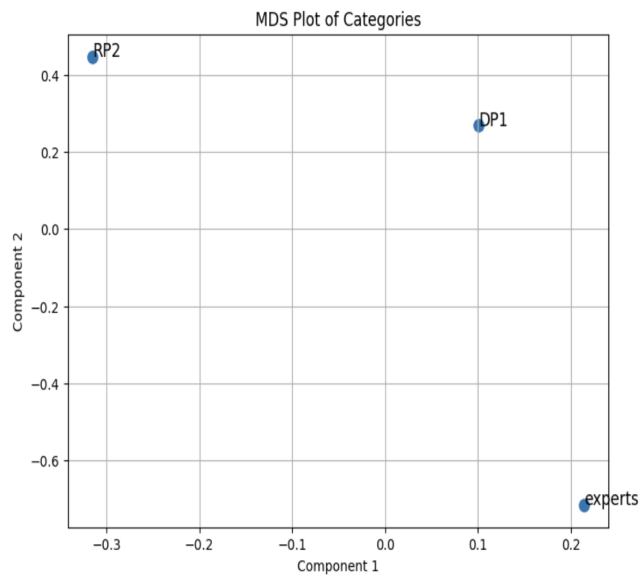

FIG 5. Multi-dimensional Scaling (MDS) plot illustrating the relative similarities between the three categories of respondents (experts, RP2 students and DP1 students).

Figure 5 represents a multi-dimensional scaling (MDS) plot showing the similarities between the three categories of respondents (experts, RP2 students and DP1 students). Multidimensional Scaling (MDS) is a statistical technique used to visualize the level of similarity or dissimilarity between sets of objects. Relative distances between the points on the plot reflect the similarities between categories on how they make problem solving decisions. Figure 5 reveals that the 'experts' category is relatively distant from both RP2 and DP1, which suggests that the pattern of responses or attributes associated with 'experts' differs substantially from those of the other two categories. RP2 and DP1, are positioned closer to each other than to experts. This proximity indicates that RP2 and DP1 share greater similarity with each other, in terms of their problem-solving decision pattern, than either share with experts. It is also apparent that DP1 is positioned closer to experts than RP2 is. This suggests that, of the two, DP1 demonstrates greater similarity to experts in problem solving decisions.

## V. Discussion

We find that teaching students a specific problem-solving framework, with the use of deliberative practice, does make them more expert-like in their problem-solving decisions compared to the use of repeated-practice (with no problem-solving framework and lots of repetition). This study contributes to the literature which emphasizes the importance of explicit instruction in problem-solving frameworks within undergraduate physics education. As established in previous research [12-14], experts approach problems not merely by applying formulas, but by strategically structuring their solution process, incorporating metacognitive reflection and adopting coherent frameworks. Our results align with these findings and extend them by demonstrating quantitatively that students who were taught to apply a structured problem-solving framework exhibited decision patterns more closely resembling those of experts than students who were not given explicit framework instruction.

This distinction underscores the pedagogical value of deliberate practice as articulated in the literature [15]. Deliberate practice—characterized by breaking down complex skills into sub-skills, providing focused and ongoing feedback, and encouraging incremental mastery—was central to the instructional approach given to the DP1 cohort. The alignment of their problem-solving decisions with those of experts highlights the importance of structured, feedback-rich practice in cultivating higher-order problem-solving skills. In contrast, repeated practice, which emphasizes rote repetition with no targeted feedback, appears less effective for achieving expert-like problem-solving skills.

Our findings support previous assertions that metacognitive activities, especially those prompting students to evaluate the reasonableness of their answers, are essential for developing true expertise [8]. The observed separation between 'experts' and students who did not engage with a structured problem-solving framework (RP2) on the MDS plot suggests that content knowledge alone is inadequate for expertise; without scaffolding, students tend to rely on surface-level cues rather than engaging deeply with underlying physical principles. Our study provides empirical support for the targeted use of problem-solving frameworks as effective pedagogical tools. Not only do these frameworks scaffold novice performance, but—as indicated by the proximity between DP1 and experts—they also help undergraduate students develop expert-like problem-solving. This suggests that integrating deliberate practice into physics instruction could help bridge the gap between novice and expert problem-solving skills.

While the findings provide strong evidence for the effectiveness of explicit problem-solving frameworks and deliberate practice pedagogies, several limitations should be acknowledged. First, the study was conducted within the context of introductory physics courses at a single institution and therefore, the generalizability of the results to other academic settings, disciplines, or educational levels are limited.

Second, the analysis relies on responses to a diagnostic instrument and self-reported strategies, which are subject to interpretation biases and might not fully capture the complex cognitive processes underlying expert problem-solving. While efforts were made to ensure consistency in coding and interpretation, some subjectivity in coding problem-solving decisions is inevitable, despite consensus procedures among expert coders. Third, the two student comparison groups (DP1 and RP2) were different student groups (albeit with similar physics backgrounds). This disparity could influence the observed distances in the MDS plot and may not account for potential confounders, such as prior knowledge and motivation. Additionally, the duration over which the interventions were implemented was relatively short. Longitudinal studies would be necessary to determine the persistence of the observed differences and whether students continue to exhibit more expert-like behaviors over time. Additionally, while the study highlights overall trends in problem-solving decisions, it does not disentangle the relative contributions of individual instructional techniques or feedback components within the broader deliberate practice approach. Further investigation, employing both quantitative and qualitative methodologies, would help clarify which aspects of the instructional interventions are most critical for developing expertise. Despite these limitations, the findings offer valuable insights into physics instruction and provide a foundation for future research aimed at optimizing problem-solving pedagogy.

## VI. Conclusion

In this study, we demonstrate that teaching students to apply an explicit problem-solving framework, coupled with deliberate practice pedagogy, contributes to the development of expert-like problem-solving behaviors in undergraduate physics courses. We show that students exposed to these targeted instructional strategies become more similar to experts in their decision-making processes than those engaging in unstructured or repetitive practice alone. These results reinforce prior literature advocating for the integration of metacognitive scaffolding and targeted feedback into science education, highlighting that deliberate practice—not mere repetition—drives meaningful skill acquisition. Given these findings, we strongly recommend that physics instructors adopt explicit problem-solving frameworks as a core component of their pedagogy and prioritize deliberate practice techniques. This approach appears to help the transition from novice to expert-like thinking. Future research should expand on these findings by exploring the long-term impact and transferability of these instructional techniques across diverse educational contexts.

## VII. Acknowledgements